# Hotspot Relaxation Dynamics in a Current-Carrying Superconductor

F. Marsili[1*], M. J. Stevens[2], A. Kozorezov[3], V. B. Verma[2], Colin Lambert[3], J. A. Stern[1], R. Horansky[2], S. Dyer[2], S. Duff[2], D. P. Pappas[2], A. Lita[2], M. D. Shaw[1], R. P. Mirin[2], and S. W. Nam[2]

[1]*Jet Propulsion Laboratory, California Institute of Technology, 4800 Oak Grove Dr., Pasadena, California 91109, USA*

[2]*National Institute of Standards and Technology, 325 Broadway, Boulder, CO 80305, USA*

[3]*Department of Physics, Lancaster University, Lancaster, UK, LA1 4YB*

[*]*corresponding author: francesco.marsili.dr@jpl.nasa.gov*

**Abstract.** We experimentally studied the dynamics of optically excited hotspots in current-carrying WSi superconducting nanowires as a function of bias current, bath temperature and excitation wavelength. We discovered that: (1) the hotspot relaxation is a factor of ~ 4 slower in WSi than in NbN; (2) the hotspot relaxation time depends on bias current, and (3) the current dependence of the hotspot relaxation time changes with temperature and wavelength. We explained all of these effects with a model based on quasi-particle recombination.

When a photon is absorbed in a superconductor, it creates a non-equilibrium region referred to as a *hotspot* [1]. The optical excitation of hotspots underpins the operation of most superconducting single photon detectors, such as microwave kinetic inductance detectors (MKIDs) [2,3], superconducting tunnel junctions (STJs) [4], and superconducting nanowire single photon detectors (SNSPDs) [5]. If hotspot dynamics were understood and controlled, many of the current limitations of these detectors would be overcome, thus enabling disruptive technologies. Here we report a combined experimental and theoretical study of hotspots excited by single photons in current-carrying WSi superconducting nanowires. We observed for the first time that: (1) the hotspot relaxation time ($t_{HS}$) depends on the current carried by the nanowires; and (2) the current dependence of $t_{HS}$ changes with bath temperature and excitation wavelength. The agreement between theory and experiment provides new insight into the quasiparticle dynamics in superconductors and the photodetection mechanism of superconducting single photon detectors.

Hotspot formation is initiated when one photon is absorbed in a thin superconducting film, creating a non-equilibrium distribution of quasiparticles (QPs). The excited QPs down-convert from higher-energy states by exchanging energy with the electron and phonon systems. During the decay, further Cooper pairs are broken, increasing the number of QPs [1,6]. Previously, the relaxation of optically excited superconductors was studied with optical and THz pump-probe techniques [7-9]. These techniques offer sub-ps time resolution, but are not sensitive enough to study the evolution of a single hotspot, and are difficult to perform below ~ 5 K. We used a different technique that combines the single-hotspot sensitivity of electrical readout with the high time resolution of ultrafast optical pump-probe spectroscopy [10,11].

We measured the dependence of $t_{HS}$ on bias current ($I_B$), bath temperature ($T_B$), and excitation wavelength ($\lambda$) using a fiber-coupled WSi SNSPD [12,13] based on ~ 5 nm thick, 130 nm-wide nanowires spaced on a 200 nm pitch, meandering over an 11 μm-diameter circular active area. The SNSPD was operated in an adiabatic demagnetization refrigerator, in the temperature range $T_B$ = 0.25 - 2 K. The source of optical excitation was a fiber-based ultrafast pulsed supercontinuum source (1200 - 1650 nm wavelength range, ~ 5 ps pulse duration, $f_{rep}$ = 36 MHz repetition rate). We selected a given excitation wavelength with one of several band-pass filters (each ~ 12 nm bandwidth) placed between the supercontinuum source and the detector. To measure the hotspot relaxation time we produced optical pulse pairs separated by a variable delay by coupling the laser to a Mach-Zehnder interferometer [10].

If an SNSPD produces an output pulse when a single photon creates a single hotspot, the detector operates in the *single-photon detection regime*. When the bias current is lowered to a point that a response pulse can be efficiently triggered only if two photons generate two overlapping hotspots [5,10,14], the SNSPD operates in the *two-photon detection regime*. We measured the hotspot relaxation time by biasing an SNSPD in the two-photon detection regime and exciting it with two successive light pulses separated by a variable time delay ($t_D$).

To isolate the bias range for two-photon detection, we coupled the filtered supercontinuum source to a bank of calibrated attenuators and then to the detector [13]. We measured the probability of detection per optical pulse ($P_{click}$) as a function of the mean photon number per pulse ($\mu$, which was determined as described in Ref. [13]) at several fixed values of $I_B$ (see section 1 of Supplemental Material). We defined $P_{click}$ as the ration between the photoresponse count rate (*PCR*) and $f_{rep}$. *PCR* was calculated as the difference between the count rate measured with the light source coupled to the detector (*CR*) and the count rate measured with the light source blanked with a shutter (background count rate, *BCR*) [13].

In the limit where only one- and two-photon detection events lead to a measureable count rate, the click probability can be written as [15]:

$$P_{\text{click}}(\mu) = 1 - e^{-\mu} \sum_{n=0}^{\infty} \frac{\mu^n}{n!} (1-\eta_1)^n (1-\eta_2)^{n(n-1)/2} \tag{1}$$

where $\eta_1$ is the single-photon system detection efficiency (which we defined as the probability that a photon coupled in the SNSPD fiber created a response pulse), $\eta_2$ is the two-photon system detection efficiency (which we defined as the probability that two photon coupled in the SNSPD fiber created a response pulse), and *n* is the number of photons per pulse coupled to the detector system. $\eta_1$ and $\eta_2$ depend on the bias current, bath temperature, and excitation wavelength. Equation (1) ignores the effect of dark counts, since *BCR* had already been subtracted from the data. In the single-photon detection regime, $\eta_1 \gg \eta_2$ and Equation (1) simplifies to $P_{\text{click}}(\mu) = 1 - \exp(-\eta_1 \cdot \mu)$. If $\eta_1 \cdot \mu \ll 1$, we obtain the familiar approximation: $P_{\text{click}}(\mu) \sim \eta_1 \cdot \mu$. If the detector operates in the two-photon detection regime (where $\eta_2 \gg \eta_1$) and $\eta_2 \cdot \mu^2 \ll 1$, Eq. (1) can be approximated by $P_{\text{click}}(\mu) \sim \eta_2 \cdot \mu^2 / 2$. Following the method described in Ref. [15] we extracted the bias dependence of $\eta_1$ and $\eta_2$ (see section 1 of Supplemental Material). For $1.9~\mu A \leq I_B \leq 3.5~\mu A$, the SNSPD operated in the two-photon-detection regime, with $\eta_1 \ll \eta_2$.

To measure the hotspot relaxation time we illuminated the detector with the pulse-pair source and measured *PCR* as a function of $t_D$ over a range of 1 ns. Figure 1 a shows *PCR* vs $t_D$ curves measured at different bias currents. The *PCR* vs $t_D$ curves had a Lorentzian shape except in the range -5 ps $\leq t_D \leq$ 5 ps, where the *PCR* exhibited oscillations due to the optical interference of overlapping pulse pairs (the region enclosed by a dashed square in Figure 1 a). Surprisingly, the *PCR* vs $t_D$ curves became broader as the bias current increased. The dependence of *PCR* on $t_D$ can be correlated with the hotspot relaxation dynamics using a simple argument [10,11], valid only if the detector operates in the two-photon detection regime and $\eta_2 \cdot \mu^2 \ll 1$. If the time delay between pulses is longer than the hotspot relaxation time ($t_D > t_{HS}$), $P_{\text{click}}$ is approximately equal to the sum of the probabilities from each pulse acting independently: $P_{\text{click}}(t_D \gg t_{HS}) \approx \eta_2 \mu^2$. Close to zero delay ($t_D \ll t_{HS}$), the two pulses do not overlap but act as a single pulse with twice the mean photon number, so the click probability doubles: $P_{\text{click}}(t_D \ll t_{HS}) \approx \eta_2 (2\mu)^2 / 2 = 2\eta_2 \mu^2$. As a result, we expect $P_{\text{click}}(t_D \ll t_{HS}) / P_{\text{click}}(t_D \gg t_{HS}) \approx 2$, which is in agreement with the data shown in Figure 1 a.

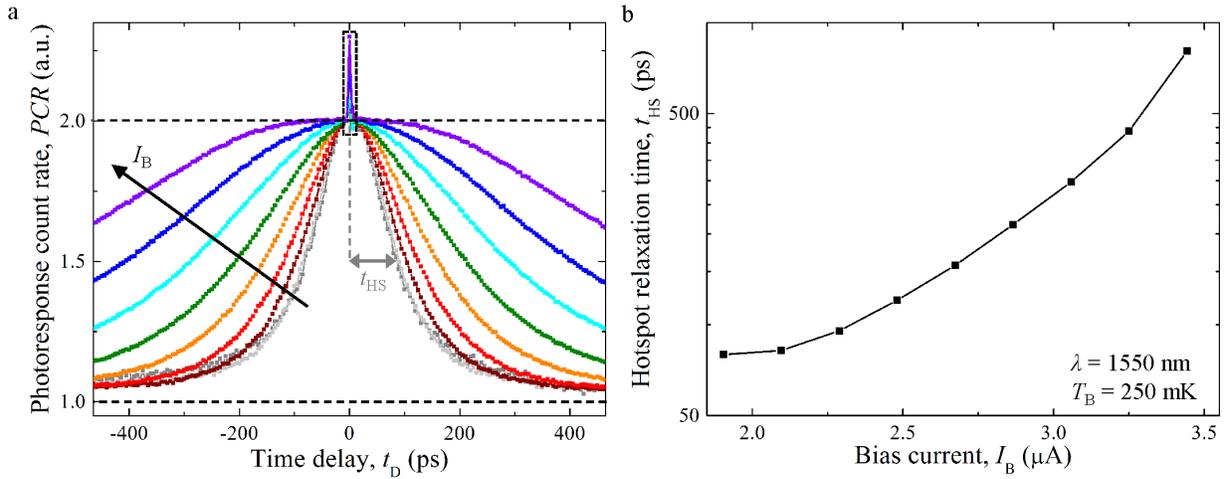

**Figure 1. a.** Normalized *PCR* vs $t_D$ curves measured with the detector operating in two-photon detection regime at $I_B$ = 1.9 µA (dark gray); 2.1 µA (light gray); 2.3 µA (dark red); 2.5 µA (red); 2.7 µA (orange); 2.9 µA (green); 3.1 µA (cyan); 3.3 µA (blue); 3.5 µA (violet). The black arrow indicates the direction of increasing $I_B$. The dark gray arrow indicates the half width at half maximum (HWHM) of the *PCR* vs $t_D$ curve measured at $I_B$ = 1.9 µA. Each of the *PCR* vs $t_D$ curves was normalized so that the maximum value of the corresponding Lorentzian fit curve was 2. The supercontinuum source was attenuated so that probability of detection per optical pulse (or click probability, Pclick) $P_{\text{click}}$ < 10 % in each individual pulse. **b.** $t_{HS}$ vs $I_B$ curve extracted from fits to the data in panel a. These measurements were performed at $\lambda$ = 1550 nm, and $T_B$ = 0.25 K. The switching current of the device, which is defined as the maximum current the device can be biased at without switching to the normal, non-superconducting state, was $I_{SW}$ = 8.8 µA.

To extract $t_{HS}$ from the *PCR* vs $t_D$ data, we fit the experimental curves with Lorentzians, ignoring data in the range of the optical interference. We defined $t_{HS}$ as the half width at half maximum (HWHM) of the fitting curves. Figure 1 b shows the $t_{HS}$ vs $I_B$ curve extracted from the data shown in Figure 1 a. When $I_B$ was increased from $I_B$ = 1.9 µA to $I_B$ = 3.5 µA, $t_{HS}$ increased by one order of magnitude from $t_{HS}$ ~ 80 ps to $t_{HS}$ ~ 800 ps. To our knowledge, this effect had not been observed before. Our discovery suggests that the quasiparticle relaxation time of MKIDs based on disordered materials [3,16] may be increased by DC-biasing the MKID inductor, which would increase the MKID sensitivity. Furthermore, the shortest hotspot relaxation time measured in our WSi SNSPD is a factor of ~ 4 longer than that measured with NbN (15-30 ps [1,10,11,17]), indicating a significant difference in material properties that was not well understood and not previously predicted. The longer $t_{HS}$ of WSi may limit the maximum count rate of WSi SNSPDs to a lower value than NbN SNSPDs, due to latching [18,19] or afterpulsing [20].

To gain insight into the mechanism that caused $t_{HS}$ to increase when $I_B$ was increased, we investigated how the bias dependence of $t_{HS}$ changed when changing $T_B$ and $\lambda$. As shown in Figure 2 a, $t_{HS}$ increased when $T_B$ was increased at fixed wavelength (squares). As shown in Figure 2 b, $t_{HS}$ decreased when $\lambda$ was increased at fixed temperature (squares). The shape of the $t_{HS}$ vs $I_B$ curves measured at different temperatures and wavelengths shows a correlation with the temperature and wavelength dependence of the cutoff current $I_{co}$ [21], which represents the current above which SNSPDs operate in the single-photon detection regime (see section 2 in Supplemental Material). As shown in Figure 2 c, while the $t_{HS}$ vs $I_B$ curves measured at different temperatures and wavelengths differ significantly, the $t_{HS}$ vs $I_B / I_{co}$ curves (squares) closely follow the same trend, indicating a correlation between hotspot dynamics and device sensitivity.

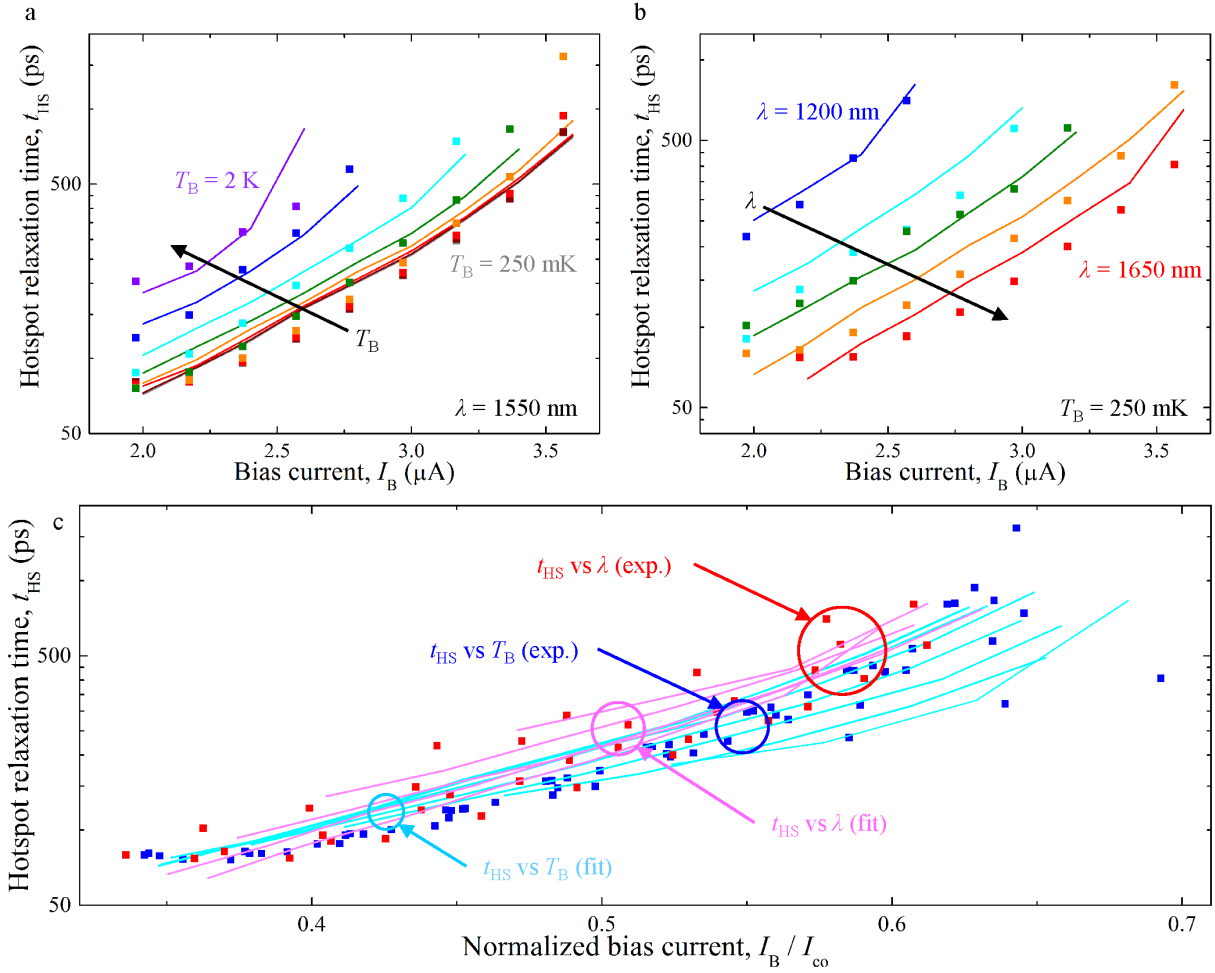

**Figure 2. a.** Experimental $t_{HS}$ vs $I_B$ curves measured at different bath temperatures (squares) and fitting curves (lines). The bath temperatures were $T_B$ = 0.25 (gray), 0.5 (dark red, overlapping with gray), 0.75 (red), 1 (orange), 1.25 (green), 1.5 (cyan), 1.75 (blue), 2 K (violet). The excitation wavelength was $\lambda$ = 1550 nm. The black arrow indicates the direction of increasing $T_B$. The values of the fitting parameters are: $\gamma$ = 0.3, indicating that non-equilibrium phonons deposit energy into the electronic system before escaping to the substrate; $\tau_0$ = 497 ps, which is commensurate to the $\tau_0$ of materials with order parameter similar to WSi [22]; $\delta$ = 325 meV$^{-1}$, indicating that only a small fraction ($\chi$ = 0.13) of the energy of the photon is deposited into the electronic system (likely due to: (1) the energy partition between QPs and non-pair-breaking phonons, and (2) the loss of athermal phonons [23]); and $\Delta T_B$ = 0.5 K, which may be due to the laser heating the sample. **b.** Experimental $t_{HS}$ vs $I_B$ curves measured at different wavelengths (squares) and fitting curves (lines). The excitation wavelength were $\lambda$ = 1200 nm (blue); 1350 nm (cyan); 1450 nm (green); 1550 nm (orange); 1650 nm (red). The bath temperature was $T_B$ = 250 mK. The black arrow indicates the direction of increasing $\lambda$. The values of the fitting parameters are: $\gamma$ = 0.3, $\tau_0$ = 439 ps, $\delta$ = 325 meV$^{-1}$. The values of the fitting parameters agree with those used to fit the data in Figure 5 a. **c.** Blue squares (cyan lines): experimental (simulated) $t_{HS}$ vs normalized bias current measured at different temperatures ($T_B$ = 0.25 – 2 K) and fixed wavelength ($\lambda$ = 1550 nm), as shown in panel a. Red squares (magenta lines): experimental (simulated) $t_{HS}$ vs normalized bias current measured at different wavelengths

($\lambda$ = 1200 -1650 nm) and fixed bath temperature ($T_B$ = 250 mK), as shown in panel b. The bias current of the curves measured at each temperature and wavelength were normalized by the corresponding cutoff current (see section 2 in Supplemental Material for the $I_{co}$ vs $T$ and $I_{co}$ vs $\lambda$ curves).

We have developed a theoretical model for the bias dependence of $t_{HS}$ in which QP recombination is the dominant hotspot relaxation mechanism and QP diffusion is ignored. Since neglecting QP diffusion is in contrast with the traditional theory of hotspot dynamics [1,24-26], our model provides new insight into the physics of non-equilibrium superconductivity. Our model quantitatively reproduces the experimentally observed decrease in $t_{HS}$ when: (1) decreasing the bias current, (2) decreasing the bath temperature, and (3) increasing the excitation wavelength. Furthermore, our model provides an estimate for the $t_{HS}$ of NbN close the experimental values [10,11,17] (see section 6 of Supplemental Material). The details of our model are discussed elsewhere [27] (see section 3 of Supplemental Material).

Our model simulates the time evolution of the QP temperature ($T_{QP}$) after the absorption of two subsequent photons under a variety of conditions (changing $t_D$, $I_B$, and $T_B$). As shown in Figure 3 a, after absorption of the first photon, $T_{QP}$ instantly increases from $T_B$ to the excitation temperature ($T_{ex}$) and then starts relaxing back to $T_B$. When the second photon is absorbed in the hotspot (after a delay time $t_{D1}$), $T_{QP}$ exceeds the critical temperature ($T_C$, which depends on the bias current [27]), so the hotspot switches to the normal state, resulting in an output pulse (or *click*). Figure 3 b shows the time evolution of $T_{QP}$ for the same conditions as in Figure 3 a, except for a longer delay time ($t_{D2} > t_{D1}$). In this case, absorption of the second photon does not cause $T_{QP}$ to exceed $T_C$ and no click is produced. We defined a cutoff temperature ($T_{co}$) as the lowest QP temperature at which absorption of the second photon causes a click ($T_{QP} = T_C$). We defined the theoretical hotspot relaxation time ($t_{HS}^t$) as the time required for $T_{QP}$ to reach $T_{co}$.

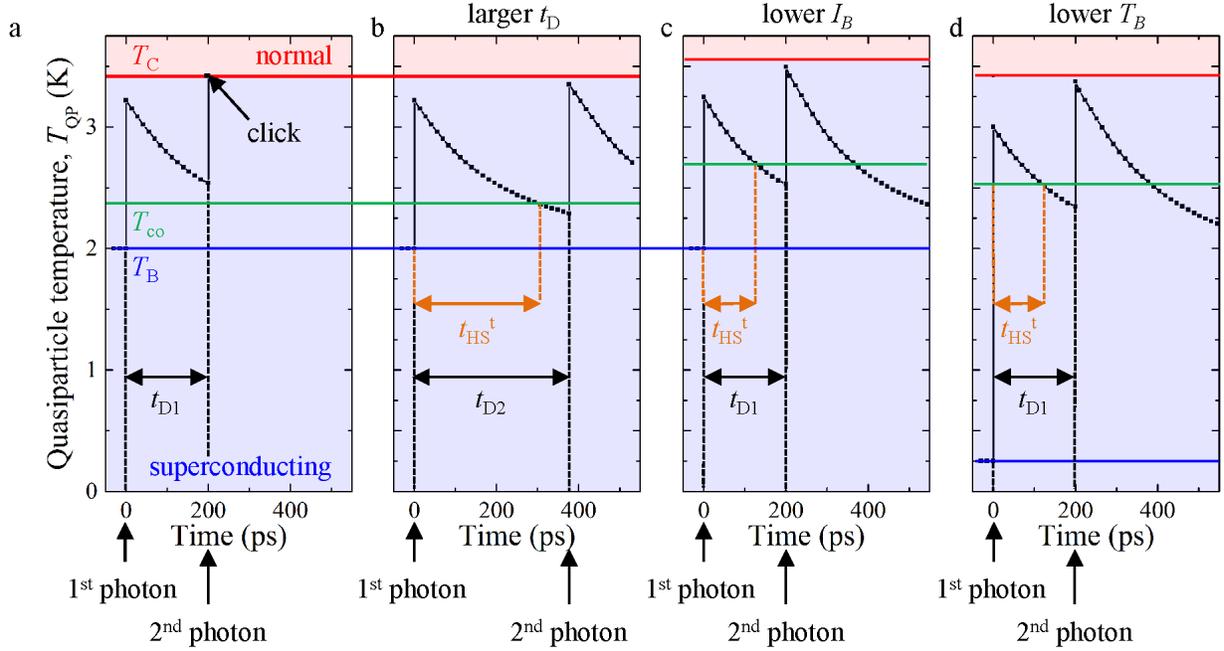

**Figure 3. a.** Simulated time evolution of $T_{QP}$ (black squares) after the absorption of two subsequent photons at 0 s and $t_{D1} = 200$ ps. The parameters of the simulation are: $T_B = 2$ K, $T_C = 3.4$ K, $I_B = 2.4$ μA, $\lambda = 1500$ nm. The blue line represents $T_B$, the red line represents $T_C$ and the green line represents $T_{co} = 2.4$ K. The range of temperatures for which the hotspot is superconducting is colored in blue; the range for which the hotspot is normal (resistive) is colored in red. The critical temperature of the nanowire at zero bias is 4.5 K. **b.** Simulated time evolution of $T_{QP}$ for the same parameters as panel a, except for a longer delay time $t_{D2} = 375$ ps. The theoretical hotspot relaxation time is $t_{HS}^t = 310$ ps. **c.** Simulated time evolution of $T_{QP}$ for the same parameters as panel a, except for a lower bias current $I_B = 2$ μA. The theoretical hotspot relaxation time decreases to $t_{HS}^t = 125$ ps. **d.** Simulated time evolution of $T_{QP}$ for the same parameters as panel a, except for a lower bath temperature $T_B = 0.25$ K. The theoretical hotspot relaxation time decreases to $t_{HS}^t = 110$ ps. The orange arrows highlight the theoretical hotspot relaxation time ($t_{HS}^t$).

We attributed the dependence of $t_{HS}$ on $I_B$ shown in Figure 1 to the increase of $T_{co}$ when $I_B$ is decreased. As shown in Figure 3 c, since $T_C$ increases when $I_B$ is decreased [27], $T_{co}$ also increases. $T_{ex}$ and the relaxation transient of $T_{QP}$ are not significantly affected by the change in bias current. Therefore, if $I_B$ is decreased, $T_{QP}$ relaxes from $T_{ex}$ to $T_{co}$ in a shorter time, in agreement with the results in Figure 1.

The dependence of $t_{HS}$ on $T_B$ shown in Figure 2 a can be attributed to the temperature dependence of the relaxation rate of $T_{QP}$. As shown in Figure 3 d, while $T_B$ does not affect $T_{co}$, $T_B$ does affect $T_{ex}$ and the relaxation transient of $T_{QP}$. Our model predicts that: (1) at lower $T_B$, $T_{ex}$ is lower, and (2) the relaxation rate of $T_{QP}$ is higher when $T_{QP}$ is further away from $T_B$. Consequently, $t_{HS}^{t}$ decreases when the $T_B$ is decreased, in agreement with the experimental results in Figure 2 a.

We attributed the dependence of $t_{HS}$ on $\lambda$ shown in Figure 2 b to the dependence of $T_{ex}$ and $T_{co}$ on the photon energy. According to our model, with longer-wavelength photons: (1) the increase of $T_{QP}$ after the absorption of a photon ($T_{ex}$ - $T_B$) is smaller, and (2) $T_{co}$ is higher. At longer excitation wavelengths, $T_{ex}$ is lower and $T_{co}$ is higher, so $T_{QP}$ decreases from $T_{ex}$ to $T_{co}$ more quickly, in agreement with the experimental results shown in Figure 2 b.

The solid curves in Figure 2 a and b show fits to the experimental data (squares) calculated with our model. We fit all the experimental data in Figure 2 a using four fitting parameters: (1) the phonon bottleneck parameter $\gamma = \tau_{esc} / \tau_{ph-e}$, where $\tau_{esc}$ is the phonon escape time to the substrate and $\tau_{ph-e}$ is the phonon-electron scattering time; (2) the characteristic quasiparticle time of WSi ($\tau_0$, as defined in Ref. [22]); (3) the energy deposition factor $\delta = \chi / \varepsilon_c$, where $\chi = E_{ex} / E_\lambda$ is the photon yield, which we defined as the ratio between the energy deposited in the hotspot after the absorption of the photon ($E_{ex}$) and the photon energy ($E_\lambda$), and $\varepsilon_c$ is the energy of the condensate in the hotspot volume; and (4) a temperature offset ($\Delta T_B$), which we defined as the difference between the simulated and experimental bath temperatures. We fitted all of the experimental data in Figure 2 b using three fitting parameters: $\gamma$, $\tau_0$, and $\delta$. Our model can also accurately predict the temperature and wavelength dependence of $I_{co}$. Using the values of the fitting parameters obtained from the fits shown in Figure 2 a and b, we could reproduce the shape of the experimental $I_{co}$ vs $T_B$ and $I_{co}$ vs $\lambda$ curves (shown in section 2 of Supplemental Material) [27]. Figure 2 c shows the experimental and simulated $t_{HS}$ vs $I_B / I_{co}$ curves at different temperatures and wavelengths. The four families of curves closely follow the same trend. The correlation between the hotspot dynamics and SNSPD sensitivity is discussed in detail in Ref. [27].

In summary, we observed for the first time that the hotspot relaxation time of a superconducting nanowire can be increased by increasing the bias current. We developed a model that explains and quantitatively reproduces all the experimental data. The effect we discovered provides new insight into non-equilibrium superconductivity and has important implications for superconducting detectors.

**Acknowledgements**

Part of the research was carried out at the Jet Propulsion Laboratory, California Institute of Technology, under a contract with the National Aeronautics and Space Administration. AK and CL gratefully acknowledge financial support from the Engineering and Physical Sciences Research Council.